\documentclass[journal=ancac3,manuscript=article]{achemso}
\usepackage{achemso}
\setkeys{acs}{maxauthors = 50}
\usepackage{graphicx}
\usepackage{graphics}
\usepackage{amsmath}
\usepackage{amsfonts}
\usepackage{color}
\usepackage{caption}
\usepackage{amssymb}
\usepackage{subcaption}
\usepackage{siunitx}
\usepackage{tikz}
\usepackage[export]{adjustbox}
\usepackage{multicol}

\usepackage{soul}
\def\rmd{{\mathrm{d}}}

\author{Zhujie Li}
\affiliation[]
{Applied Theoretical Physics-Computational Physics, Physikalisches Institut, Albert-Ludwigs-Universit{\"a}t Freiburg, D-79104 Freiburg, Germany}

\author{Victor G. Ruiz}
\affiliation[]
{Research Group for Simulations of Energy Materials, Helmholtz-Zentrum Berlin, D-14109 Berlin, Germany}
\author{Matej Kandu{\v c}}
\affiliation[]
{Jo{\v z}ef Stefan Institute, SI-1000 Ljubljana, Slovenia}
\author{Joachim Dzubiella}
\email{joachim.dzubiella@physik.uni-freiburg.de}
\affiliation[]
{Applied Theoretical Physics-Computational Physics, Physikalisches Institut, Albert-Ludwigs-Universit{\"a}t Freiburg, D-79104 Freiburg, Germany}
\alsoaffiliation[]
{Research Group for Simulations of Energy Materials, Helmholtz-Zentrum Berlin, D-14109 Berlin, Germany}

\title
{Ion-specific Adsorption on Bare Gold (Au) Nanoparticles in Aqueous Solution: Double-Layer Structure and Surface Potentials}

\abbreviations{} 

\makeatother

\begin{document}
	\begin{abstract}
		
		We study the solvation and electrostatic properties of bare gold (Au) nanoparticles (NPs) of $1$--$2$~nm in size in aqueous electrolyte solutions of sodium salts of various anions with large physicochemical diversity (Cl$^-$, BF$_4$$^-$, PF$_6$$^-$, Nip$^-$(nitrophenolate), 3- and 4-valent hexacyanoferrate (HCF)) using nonpolarizable, classical molecular dynamics computer simulations. We find a substantial facet selectivity in the adsorption structure and spatial distribution of the ions at the Au-NPs: while sodium and some of the anions (e.g., Cl$^-$, HCF$^{3-}$) adsorb more at the `edgy' (100) and (110) facets of the NPs, where the water hydration structure is more disordered, other ions (e.g., BF$_4$$^-$, PF$_6$$^-$, Nip$^-$) prefer to adsorb strongly on the extended and rather flat (111) facets. In particular, Nip$^-$, which features an aromatic ring in its chemical structure, adsorbs strongly and perturbs the first water monolayer structure on the NP (111) facets substantially. Moreover, we calculate adsorptions, radially-resolved electrostatic potentials, as well as the far-field {\it effective} electrostatic surface charges and potentials by mapping the long-range decay of the calculated electrostatic potential distribution onto the standard Debye--H{\"u}ckel form. We show how the extrapolation of these values to other ionic strengths can be performed by an analytical Adsorption-Grahame relation between effective surface charge and potential.  We find for all salts negative effective surface potentials in the range from $-10$~mV for NaCl down to about $-80$~mV for NaNip, consistent with typical experimental ranges for the zeta-potential. We discuss how these values depend on the surface definition and compare them to the explicitly calculated electrostatic potentials near the NP surface, which are highly oscillatory in the $\pm 0.5$~V range.  
		
	\end{abstract}

	\section{Introduction}
	
	The research and literature on gold nanoparticles (AuNPs) has exploded in the last decade because of their versatility and usefulness in a wide range of applications, such as cancer therapy,\cite{cancer,cancer2} drug delivery,\cite{biology,delivery2,drug} plasmonic resonance,\cite{plasmon,Nguyen2019} colorimetric detection of toxins\cite{toxin}, or nanocatalysis~\cite{enzyme,restructuring,Wu2012a,Herves,catalysis,Roa2017a}. Evidently, AuNPs of nanometer size have unique optical, electronic, and structural properties, which are very different to those of the bulk state or the single atom.~\cite{synthesis,defect,Bond2012}  The synthesis of AuNPs  is primarily done by chemical methods, such as chemical reduction, however, only in the presence of ligands. Hence, it is still difficult to distinguish whether the unique  properties of AuNPs originate from the cores of the nanoparticles, bare surface structures, or from synergies with the adsorbates on their surfaces.  In particular, electrocatalytic processes in the liquid phase are influenced strongly by the interaction of the surface with the solvent and electrolytes,~\cite{synthesis,defect} yet the direct investigation of the solvation of nanoparticles is difficult.~\cite{Novelli2019}
	
	Recently, laser-fragmentation (or laser-ablation) in liquids has become a viable route to synthesize surfactant-free/ligand-free AuNPs in water.~\cite{Sylvestre,Rehbock,Ziefuss2019,DeAndaVilla2019} This technique makes it possible to study ligand and bare AuNP effects separately and control functionalization using post-synthesis chemistry, important for biological applications and catalysis in aqueous systems.  It was demonstrated, for example,  that during the laser-fragmentation process the type of halide anion that is present can be used to control the particle size distribution of the AuNP, attributed to the adsorption of the specific anions,\cite{Ziefuss2019} probably involving selective binding on crystal facets.\cite{Ghosh2018} In addition, it was shown that colloidal stabilization and dispersion of the laser-ablated NPs also depends critically on the type of anion in solution, following Hofmeister-like behaviors.~\cite{Horinek2008,Schwierz2010,Schwierz2013}  It was conjectured that this was also due to ion-specific adsorption at the gold/water interface, leading to ion-specific repulsive electrostatic interactions between the gold particles.~\cite{Merk2014b}  In fact, \textit{in situ} studies have shown that halides can form ordered adlayer structures on single-crystal electrode surfaces of Au(111) and other fcc metals in contact with aqueous solutions.~\cite{Magnussen2002} Metal nanoparticles can be also stabilized in complex ionic liquid solutions,  e.g., with anions tetrafluoroborate (BF$_4$$^-$) and hexafluorophosphate (PF$_6$$^-$).~\cite{pensado2011solvation,Podgorsek2013a} Moreover, in catalytic applications with AuNPs in the aqueous phase the adsorption of complex anionic reactants such as ferricyanide (HCF$^{3-}$) or nitrophenolate (Nip$^-$) is a key step in the surface reaction process,~\cite{Herves,Roa2017a}  and can be tuned by the electrostatic charge and surface potential of the nanoparticle.~\cite{roy}

	All these studies show that anionic adsorption on ligand-free AuNPs is of fundamental importance with far reaching, electrostatically-mediated consequences for applications. In particular, the highly faceted nature of the small AuNPs, which seems responsible for their high catalytic activity,\cite{restructuring,defect} may affect the surface adsorption mechanisms of the anions in a substantial way due to the proportion of atoms of low coordination number occurring at edges and corners, which increases as the size of the AuNP decreases. On the other hand, all these local structural effects define differences in the total ionic adsorptions and with that define the far-field electrostatic signature (e.g., effective charge for the electrostatic double-layer interactions in solution), important for colloidal stabilization.~\cite{Hunter} Detailed molecular-level effects are still not well understood due to the small scales involved, large fluctuations on the nanoscales, and the complex solvation processes, in particular the competition between bulk solvation and the adsorption at the interfaces. 
	
	Access to microscopic structure and adsorption mechanisms in the liquid phase is provided in principle by molecular dynamics (MD) computer simulations. MD simulations have been applied, for instance, to study adsorption of water on bare~\cite{Li2015b} or ligand-coated NPs,~\cite{Yang2015a}  peptides in water on bare AuNPs,~\cite{Monti2016a, Shao2018b, Monti2018b, Tang2019,Samieegohar2019} surfactants and halides in water on planar gold,~\cite{Meena2016b,Geada2018a} sodium citrate in water on bare AuNPs,~\cite{Perfilieva2019a} and ionic liquids around NPs~\cite{pensado2011solvation,Podgorsek2013a}. However, despite the importance of anionic adsorption and resulting electrostatics for colloidal stabilization~\cite{Merk2014b,Ziefuss2019} and the catalytic activity~\cite{Herves} of AuNPs, no simulation study has so far systematically addressed the relation between anionic adsorption structures and the resulting electrostatic surface potentials. 
	
	The aim of this work is thus to study the adsorption of anions -- representatively chosen with respect to the above applications but also varying in shape and valency --  to bare AuNPs of various sizes ($\simeq 1$--$2$~nm in radius) in aqueous solution. The foci of our study are set on i) the adsorption structure of water and ions in radial distance from the NP as well as their facet selectivity, ii) the ion-specific electrostatic potential distribution close to the NP-water interface, which is important for catalytic application, and iii) the effective far-field (Debye--H{\"u}ckel) electrostatic potential, which is relevant in larger scale studies, as in experiments probing electrophoretic processes or colloidal stability.~\cite{Hunter}  We study various anions in this work, ranging from atomic to molecular. We focus on bare AuNPs without surface oxidation or ligand functionalization in order to probe the effects of most basic features such as geometry and facet effects on anionic adsorption and electrostatics. Higher complexity, refined atomistic model (including force fields), and more `chemistry' (e.g., various surface terminations) shall be included in future studies, as we comment in the final concluding section.  
	
	\section{Simulation and analysis methods }
	
	\begin{figure}[htb]
		\centering
		\includegraphics[width=1\linewidth]{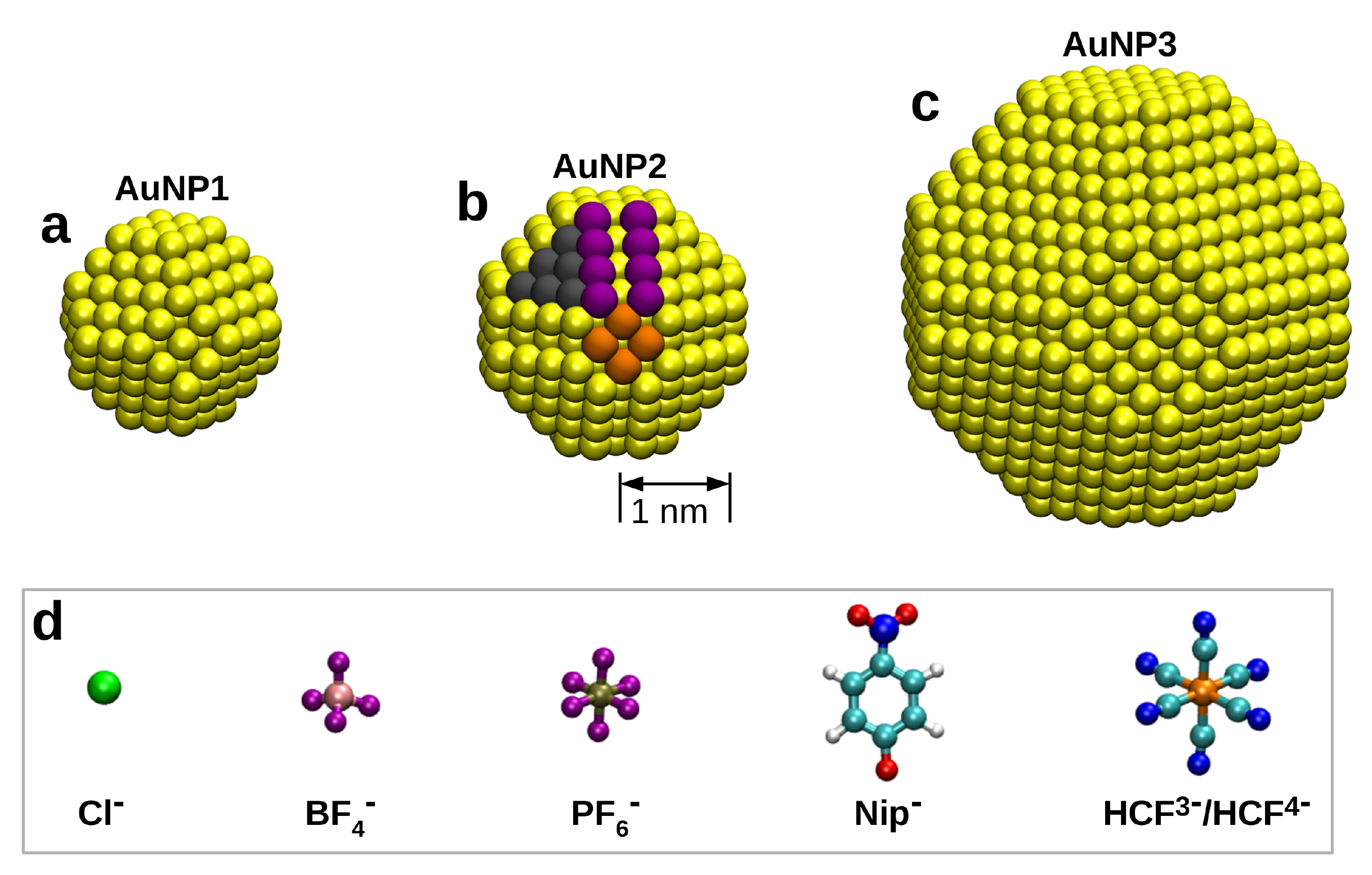} 
		\caption{a)-c) Snapshots of gold nanoparticles, AuNP1, AuNP2, AuNP3, respectively. The NP bare radii from left to right are $R_{\rm NP}=0.82, 1.02, 1.84$~nm, respectively, which are defined as the shortest distance between the center of AuNP and the (100) facets. The color code in b) stands for the representative (100) facet (orange), (111) facet (gray), and (110) facet (purple). d) Anions studied in our simulations: Cl$^-$, BF$_4$$^-$, PF$_6$$^-$, Nip$^-$, HCF$^{3-}$/HCF$^{4-}$ (from left to right). Na$^+$ ions were used as the cations (counterions) for all the simulations.}
		\label{fig:box}
	\end{figure}
	\subsection{Models and Force Fields.} 
	The atomistic model of gold (Au) nanoparticles (NPs) was built via the Wulff construction~\cite{wul1901frage}, in which the surface energies were taken from DFT calculations~\cite{Almora-Barrios2014a}. Three NPs of different sizes, namely, AuNP1, AuNP2, AuNP3, were constructed based on the truncated-octahedron morphology with bare radii of $R_{\rm NP}=0.82, 1.02, 1.84$~nm, respectively (Figure~\ref{fig:box}a-c). The bare radius is defined as the shortest distance between the center of the AuNP and the plane spanned from the centers of atoms in the (100) surfaces. The smallest NP, AuNP1, only features (100) and (111) surfaces, while the other two particles have additionally (110) facets, cf.~Figure~1a-c. Various anions (Figure~\ref{fig:box}d) were studied in this work, including chloride (Cl$^-$), as a representative of a halide ion, tetrafluoroborate (BF$_4$$^-$) and hexafluorophosphate (PF$_6$$^-$), as representatives of complex (ionic liquid) ions,~\cite{pensado2011solvation,Podgorsek2013a} and nitrophenolate (Nip$^-$), ferricyanide [Fe(CN)$_6$]$^{3-}$ (HCF(III) or HCF$^{3-}$) and ferrocyanide [Fe(CN)$_6$]$^{4-}$ (HCF(II) or HCF$^{4-}$), as representative for catalytic ions~\cite{Herves}. Sodium ions (Na$^+$) were used as the cations (counterions) for all the solutions.
	
	The initial configuration was constructed with the PACKMOL package~\cite{Martinez2009} in a cubic cell with the AuNP placed in the center. The simulation cell was chosen large enough in order to minimize finite size effects as much as possible. The number of the sodium cations was set to yield bulk concentrations (number density) between 100--130~mM for all studied electrolyte solutions, resulting in (equilibrium) bulk ionic strength between 100--300~mM. Simulation details of the simulated systems are given in Table~\ref{tbl:cell}.                  
	
	Nonpolarizable force field parameters developed by Heinz and co-workers were used for the Au atoms~\cite{Heinz2008b}. Force field parameters for Na$^+$ and Cl$^-$ were taken from the recent work,~\cite{Geada2018a} while those for BF$_4$$^-$ and PF$_6$$^-$ were taken from Lopes and Padua~\cite{CanongiaLopes2006a}. For the Nip$^-$ anion, the OPLS-based force field was used with partial charge parametrization ``OPLS/QM1'' from Kandu{\v c} \textit{et al.}~\cite{Kanduc2017c}. For the HCF anions, force field parameters were taken from Prampolini and co-workers~\cite{Prampolini2014a}. The SPC/E model was used for the water molecules~\cite{Berendsen1987}. 
	
	\subsection{Simulations.} 
	All atomistic MD simulations were performed using the LAMMPS package~\cite{Plimpton1995}. The simulations were conducted with an integration time step of 1 fs at 298.15 K. Each of the systems was first equilibrated for 5--10 ns in the NPT ensemble using the Parrinello--Rahman barostat to maintain a pressure of 1 bar with a time constant of 1 ps, in which the cell size was well converged. For better post-analysis with a constant volume, the systems were then subjected to the NVT ensemble under the Nos{\'e}-Hoover thermostat with a time constant of 1 ps for 200--300 ns production runs where the data were collected. Such long simulation times were essential to have good statistics, especially for sampling the electrostatic properties. Au atoms were not frozen during our simulations. The fluctuations of Au atom positions during production simulations appear around their corresponding initial positions (Figure~S1 in SI), confirming the stability of AuNP and the structural robustness of the chosen nonpolarizable force field.     
	
	\begin{table}
		\centering
		\caption{Parameters of simulated systems for three studied nanoparticles AuNP1, AuNP2, and AuNP3, cf. Fig.~1. $R_{\rm NP}$ is the bare radius of gold nanoparticle (see text), $R_{\rm{GDS}}$ is the radius of the  (spherical) Gibbs dividing surface, eq.~(\ref{eq:GDS}), and $R_{\rm eff}$ is the effective radius defined by the first minimum in the radial distribution function of water, cf. Fig.~\ref{fig:rdf}. $N_{\alpha}$ ($\alpha$ = Au atoms, water molecules, cations, or anions) is the number of species $\alpha$, $I$ is bulk ionic strength, $L_{\rm cell}$ is the equilibrated simulation cell length.} 
		\label{tbl:cell} %
		\scalebox{1}{%
			\begin{tabular}{ccccccccccc}
				\hline 
				\rule{0pt}{4ex}NPs&$R_{\rm{NP}}$&$R_{\rm{GDS}}$&$R_{\rm{eff}}$&$N_{\rm{Au}}$&$N_{\rm{water}}$&Salt&$N_{\rm{cation}}$&$N_{\rm anion}$&$I$&$L_{\rm cell}$\\
				\rule{0pt}{1ex}   &[$\rm{nm}$]  &[$\rm{nm}$]  &[$\rm{nm}$]  &  &  & &  &  &[$\rm{mM}$]&[$\rm{nm}$]\\           
				\hline 
				\rule{0pt}{3ex} AuNP1&0.82&0.88&1.26&201 &9141 &NaCl&22 &22 &130(5)&6.53 \\
				\rule{0pt}{4ex} AuNP2&1.02&1.09&1.47&369 &17858&NaCl&43 &43 &131(1)&8.16 \\
				\rule{0pt}{4ex} AuNP2&1.02&1.09&1.47&369 &17858&NaBF$_4$&43 &43 &119(2)&8.18 \\
				\rule{0pt}{4ex} AuNP2&1.02&1.09&1.47&369 &17858&NaPF$_6$&43 &43 &113(1)&8.19 \\
				\rule{0pt}{4ex} AuNP2&1.02&1.09&1.47&369 &17858&NaNip&60 &60 &93(3)&8.21 \\
				\rule{0pt}{4ex} AuNP2&1.02&1.09&1.47&369 &17858&Na$_3$HCF&45 &15 &249(5)&8.18 \\
				\rule{0pt}{4ex} AuNP2&1.02&1.09&1.47&369 &17858&Na$_4$HCF&40 &10 &287(13)&8.17 \\
				\rule{0pt}{4ex} AuNP3&1.84&1.91&2.35&1865&28985&NaCl&70 &70 &129(1)&9.66 \\
				[1ex]\hline 
			\end{tabular}\quad }
	\end{table}
	\subsection{Analysis Methods.}
	\subsubsection{Ionic strength}
	The ionic strength $I$ used in this work, which measures the concentration of ions in solutions, is defined as
	\begin{equation}
	I=\frac{1}{2} \sum_{i} \rho_i^0 z_i^{2}
	\label{eq:I}
	\end{equation}
	where the sum runs over all ion species $i$,  $\rho_i^0$ is the bulk density of ion species $i$, and $z_i$ is the charge number (valency) of that ion. Note that the values shown in Tables~\ref{tbl:cell} and~\ref{tbl:sum} are the bulk ionic strengths as output from the simulations, in which the ``bulk" is defined as the region beyond the radial distance $r={L_{\rm cell}}/{2}-0.5$~nm, where $L_{\rm cell}$ is the length of the cubic simulation cell. The ionic strength is a consistent quantity (for both monovalent and trivalent salts) to characterize all solutions studied in present work and thus it was used here rather than the concentration. More importantly, the ionic strength plays a crucial role in the Debye--H{\"u}ckel theory as will be discussed in the following.

	\subsubsection{Excess adsorption, Gibbs dividing radius, and coordination number}
	The average excess number (over water) of adsorbed species $i$ on the AuNP can be obtained by~\cite{Horinek2011} 
	\begin{subequations}
		\begin{align}
		N^{\rm ex}_{i} &= \int_{0}^{R_{\rm{GDS}}}\rho_i(r)4\pi r^2\rmd r + \int_{R_{\rm{GDS}}}^{\infty}[\rho_{i}(r)-\rho^0_i]4\pi r^2\rmd r \\
		&= \int_{0}^{\infty}[\rho_{i}(r)-\rho^0_{i}]4\pi r^2\rmd r + \frac{4}{3}\pi \rho^0_{i} R_{\rm{GDS}}^3
		\end{align}
		\label{eq:ex}
	\end{subequations}
	where $\rho_i(r)$ is the radial number density of species $i$ and $R_{\rm{GDS}}$ is the radius of the spherical Gibbs dividing surface (GDS). The GDS is thermodynamically defined such that the corresponding water adsorption vanishes (\textit{i.e.}, $N^{\rm ex}_{\rm w}$ = 0).~\cite{Barrat2003a}
	Therefore, assuming spherical symmetry,  the cubed Gibbs dividing radius, $R_{\rm{GDS}}$, is given by  
	\begin{equation}
	R_{\rm{GDS}}^3 = 3\int_{0}^{\infty}\left[1-\frac{\rho_{\rm w}(r)}{\rho_{\rm w}^0}\right]r^2\rmd r.
	\label{eq:GDS}
	\end{equation}
	The calculated values for $R_{\rm GDS}$ are summarized in Table~\ref{tbl:cell} for the three studied AuNPs. The GDS radius obtained by eq.~(\ref{eq:GDS}) essentially represents an effective size of the nanoparticle assuming it as a sphere. As can be seen in Table~\ref{tbl:cell}, $R_{\rm{GDS}}$ is by about 0.07 nm larger than the corresponding $R_{\rm NP}$ value for all the nanoparticles. This difference comes about because the radius of the Au atom is not accounted for in $R_{\rm NP}$ as well as the fact that the $R_{\rm{GDS}}$ results from the angular average of the octahedron geometry, which is always larger than the shortest distance from the NP center to plane of the (100) surface. Thus, $R_{\rm{GDS}}$ can be used as one possible and reasonable definition of the AuNP/solvent interface location. 
	
	In practice, due to the finite size of the simulation box, a hybrid scheme was employed to calculate $N_{\pm}^{\rm ex}$ accurately including also the long range decay of the number density profiles to infinity. In this approach, a density profile obtained from simulations was evaluated by eq.~(2) up to a cutoff distance R$_{\rm{cut}} =4$~nm (for AuNP2 system), then the analytical Debye--H\"{u}ckel ion density profile, $\rho_{\pm}(r) = (\rho^0/z_{\pm}) \exp[-z_{\pm}\beta e_0 \phi^{\rm{DH}}(r)]$, was integrated from $R_{\rm{cut}}$ to infinity.
	
	The coordination number of ion species $i$ around the nanoparticle up to a radial distance $r$ is given by 
	\begin{equation}
	N_{i}(r) = \int_{0}^{r} \rho_{i}(r') 4\pi r'^2 {\rm d}r'.
	\end{equation}
	This can be used to obtain the accumulated charge in a sphere of radius $r$, as
	\begin{equation}
	Z(r) = \sum_i z_i N_i(r). 
	\label{eq:acc}
	\end{equation}
	
	\subsubsection{Electrostatic potential and effective surface potential (far field)}
	\label{sec:DH}
	The radial electrostatic potential $\phi(r)$ can be directly evaluated from the radial charge density distributions according to Poisson's equation
	\begin{equation}
	\nabla^{2} \phi(r)=-\frac{1}{\epsilon_{0} \epsilon_{\rm{r}}}\sum_{j} z_{j} \rho_{j}(r),
	\label{eq:Poisson}
	\end{equation}
	where the sum runs over all atom species $j$ ($j$=O, H, Na$^+$, ...), $\rho_{j}(r)$ and $z_{j}$ are the radial number density distribution and the {\it partial charge} of atom species $j$, respectively. $\epsilon_{0}$ is the vacuum permittivity and $\epsilon_{\rm{r}}$ is the relative permittivity of the medium.  The latter is taken to be $\epsilon_{\rm{r}}=1$ (i.e., vacuum) for a full explicit treatment of all the charges in the system, i.e., including the solvent (water) partial charges of O and H. Alternatively, we will also use the value of $\epsilon_{\rm{r}}$ = 71 for SPC/E water~\cite{Kusalik1994a} to calculate electric fields and potentials in implicit-solvent approach, where only the non-aqueous charge contributions enter Poisson's equation (i.e., without the O and H partial charges of water molecules). This will be useful for the evaluation of the electrostatic potential in the far field where water only has homogeneous screening effects.  
	
	The potential $\phi(r)$ is obtained by integrating eq.~(\ref{eq:Poisson}) twice. After the first integration, we obtain the electric field as a function of radial distance 
	\begin{equation}
	E (r) = \frac{1}{\epsilon_{0} \epsilon_{\rm{r}}{r}^2}\int_{0}^{r} \rho_q(r^{\prime}){r^{\prime}}^2\rmd r^{\prime}
	\label{eq:field}
	\end{equation}
	where $\rho_q (r) = \sum_{j} z_{j} \rho_{j}(r)$ is the radial charge density in spherical coordinates with no angular dependence. The second integration leads to the electrostatic potential around the particle, 
	\begin{equation}
	\phi (r) = -\int_{0}^{r} E (r^{\prime})\rmd r^{\prime},
	\label{eq:phi}
	\end{equation}
	which fulfills the boundary condition $\phi (\infty)$ = 0. In order to obtain the surface potential we can read off the potential at a predefined radius, e.g., $R_{\rm GDS}$ or some other effective radius $R_{\rm eff}$ (to be defined below), which we will abbreviate as $\phi_{\rm GDS}^{\rm MD} = \phi(R_{\rm GDS})$ and $\phi_{\rm eff}^{\rm MD} = \phi(R_{\rm eff})$, respectively.  
	
	In order to gain insight into the effective properties seen in the far field~\cite{Xu2017c,Nikam2018}, \textit{i.e.}, effective surface charges and potentials as measured in electrophoretic measurements or derived from scattering experiments, we consider the Debye--H{\"u}ckel (DH) model for the radial electrostatic potential around a sphere~\cite{Debye1923}
	\begin{equation}
	e_0 \beta \phi^{\rm{DH}}(r)=Z_{\rm{eff}} l_{\rm{B}} \frac{\rm{e}^{\kappa R_{\rm{eff}}}}{1+\kappa R_{\rm{eff}}}\frac{\rm{e}^{-\kappa r}}{r},
	\label{eq:DH}
	\end{equation}
	where $R_{\rm{eff}}$ is the effective radius and $Z_{\rm{eff}}$ the charge valency of the particle, $e_0$ is the elementary charge, $\beta = 1/k_{\rm{B}}T$ is the inverse thermal energy, $l_{\rm{B}} = \beta e_0^2/4\pi \epsilon_{0} \epsilon_{\rm{r}}$ is the Bjerrum length of the medium, and $\kappa = ({8\pi l_{\rm{B}}I})^{1/2}$ is the inverse Debye length ($\lambda_{\rm{D}} = \kappa^{-1}$), which is proportional to the square root of the bulk ionic strength $I$, defined by eq.~(\ref{eq:I}).  
	
	There is no exact definition of $R_{\rm{eff}}$. Usually, the effective surface potential is obtained by electrophoretic experiments where the surface potential is defined by the zeta-potential, read-off where the water sticks to the solid surface.~\cite{Hunter} Hence, $R_{\rm{eff}}$ is in most cases assumed to be the location of the shear plane. It was argued that the location of the zeta-potential is an atomic diameter away from the surface where the diffusive-double layer (and the Debye--H\"uckel regime) starts.~\cite{Levin:shear}
	What comes closest to this notion in our case is the radius defined by the position between the first and second adsorbing water layers. Therefore, we define the effective surface radius $R_{\rm{eff}}$ for the DH mapping by the first minimum of the water radial distribution function (see Figure~2 later). Indeed, we will see that for the AuNPs we observe relatively dense and ordered first water layer where water molecules are adsorbed flat on the surface, consistent with previous studies.~\cite{Chang2008a,Geada2018a,Clabaut2020}  The values of $R_{\rm eff}$ of the studied cases are given in Table~1. We find that they do not change with addition of (specific) salt.We can also see that $R_{\rm{eff}}$ is consistently larger than the Gibbs dividing radius $R_{\rm{GDS}}$ by around 0.4 nm. Clearly, the observed difference roughly corresponds to the size of the water molecule, which reflects the width of the first water layer absorbed into $R_{\rm{eff}}$ but not into $R_{\rm{GDS}}$.
	
	Now, by multiplying by $r$ and taking the logarithm, eq.~(\ref{eq:DH}) can be rewritten as~\cite{Kalcher2009c,xu2017charged,Nikam2018}
	\begin{equation}
	\ln \left|r e_0 \beta \phi^{\rm{DH}}(r)\right|=\ln \left|Z_{\rm{eff}} l_{\rm{B}} \frac{\rm{e}^{\kappa R_{\rm{eff}}}}{1+\kappa R_{\rm{eff}}}\right|-\kappa r
	\label{eq:fit}
	\end{equation}
	where the right-hand side shows a linear function of $r$ with a negative slope of the inverse Debye length $\kappa$. By taking eq.~(\ref{eq:fit}) and mapping it directly on the far-field decay of the calculated electrostatic potential in the simulation (\textit{i.e.}, $\phi(r)$ in eq.~(\ref{eq:phi})), $\kappa$ in the simulation and the prefactor $Z_{\rm{eff}} l_{\rm{B}} ({\rm{e}^{\kappa R_{\rm{eff}}}})/({1+\kappa R_{\rm{eff}}})$ can be obtained by fitting of the left-hand side of eq.~(\ref{eq:fit}) to a linear function for the simulation data. Note that the prefactor is a function of two free parameters, the effective charge $Z_{\rm{eff}}$ and the effective radius $R_{\rm{eff}}$.  Thus, $Z_{\rm{eff}}$ can be determined uniquely only for a given radius $R_{\rm{eff}}$, which we defined above. Finally, the effective DH potential $~\phi_{\rm{eff}}^{\rm{DH}}$ at the given effective radius $R_{\rm{eff}}$ is obtained by eq.~(\ref{eq:DH}) as $~\phi_{\rm{eff}}^{\rm{DH}}=\phi^{\rm DH}(R_{\rm{eff}})$. Note, in the following we express $~\phi_{\rm{eff}}^{\rm{DH}}$ also as $~\phi_0$ for simplicity.  This  recipe to obtain $Z_{\rm eff}$ is essentially following the Alexander prescription applied in literature typically to describe colloidal charge renormalization.~\cite{alexander1984charge,Trizac2002,bocquet2002effective,Levin2004,Nikam2018}
	
	\subsubsection{Adsorption-Grahame equation}
	For interpretation and extrapolation of our results, it is useful to relate the ionic surface adsorption to the emerging surface potential. For this, we consider the charge density of a spherical particle of radius $R_{\rm{eff}}$, given by
	\begin{equation}
	\sigma = \frac{Z_{\rm{eff}} e_0}{4\pi R_{\rm{eff}}^2}
	\label{eq:sigma}
	\end{equation}  
	Combined with eq.~(\ref{eq:DH}), it can be rewritten in the framework of DH theory as
	\begin{equation}
	\sigma = \epsilon_{0} \epsilon_{\rm{r}} \phi_0 \left(\kappa + \frac{1}{R_{\rm{eff}}}\right)
	\label{eq:Grahame}
	\end{equation}  
	This is the linearized Grahame equation for a spherical particle, which relates the surface charge and surface potential and thus defines the double layer capacitance in its simplest definition.~\cite{Hunter} 
	In first approximation, the adsorption isotherm for the ions follows the Boltzmann factor~\cite{Schworer2018b} 
	\begin{equation}
	\rho_{\pm}(R_{\rm{eff}}) = \frac{\rho^0}{|z_{\pm}|} \exp(-\beta\varepsilon_{\pm} - z_{\pm} e_0\beta \phi_0) 
	\label{eq:rho_delta}
	\end{equation}  
	where $\rho_{\pm}(R_{\rm{eff}})$ are the cation and anion number densities, respectively, at the surface $R_{\rm{eff}}$ in a binding shell between $R_{\rm eff}-\delta$ and $R_{\rm eff}$ of a small width $\delta$. Here, $z_{\pm}$ and $\varepsilon_{\pm}$ are the valency and the binding affinity for the cation and anion, respectively. Assuming $ e_0\beta \phi_0 \ll 1$, which is the case in the DH regime,  eq.~(\ref{eq:rho_delta}) can be linearly expanded as a function of $\phi_0$. Also, the net surface charge density $\sigma$ can be expressed by the number densities at the surface as
	\begin{equation}
	\sigma = e_0 [z_+ \rho_+(R_{\rm{eff}}) + z_- \rho_-(R_{\rm{eff}})]\delta.
	\label{eq:sigma0} 
	\end{equation}
	Combined with the Grahame equation~(\ref{eq:Grahame}) and the linearized form of eq.~(\ref{eq:rho_delta}), the net charge density can be eliminated, and the surface potential can be finally expressed as a function of ionic adsorptions, via, what we call the {\it Adsorption-Grahame equation}, 
	\begin{equation}
	e_0\beta\phi_0 = \frac{4\pi R_{\rm{eff}} l_{\rm{B}} \Delta\Gamma}{ 1 + \kappa R_{\rm{eff}} + 4\pi R_{\rm{eff}} l_{\textrm{B}} \Gamma_\textrm{tot}}
	\label{eq:surf-pot}
	\end{equation} 
	where $\Delta\Gamma = \delta\rho_0(e^{-\beta\varepsilon_+} - e^{-\beta\varepsilon_-})$ and $\Gamma_\textrm{tot} = \delta\rho_0(z_+ e^{-\beta\varepsilon_+} - z_- e^{-\beta\varepsilon_-})$ are referred to as the relative adsorption and total charge adsorption on the nanoparticle surface, respectively. As we can see from eq.~(\ref{eq:surf-pot}), for a fully symmetric adsorption the relative adsorption vanishes and the effective surface potential is zero.  By comparing this relation to our simulation results, at least for weakly adsorbing ions for which the surface potentials are low and the expression should be valid, we can better interpret and extrapolate our findings. Considering that the adsorption of Na$^+$ cations is lower than that of anions, the cation binding affinity ($\varepsilon_+$) was set to 0 as a reference.  For the binding shell we assume a width of $\delta = 0.1$~nm, which is of the order of the width of the first peak in the RDFs of the anions (see Fig.~2). Therefore, the anion binding affinity ($\varepsilon_-$) is the only fitting parameter in eq.~(15) and can be uniquely determined by a given effective surface potential at a single concentration. (Note, the exact choice of $\delta$ has only small influence on the results due to logarithmic corrections). The binding affinities for Cl$^-$, BF$_4$$^-$, and PF$_6$$^-$ were determined and used to extrapolate the effective potential as a function of ionic strength. Note that according to eqs.~(\ref{eq:sigma}) and  (\ref{eq:sigma0}), the effective surface charge should be in principle given by the accumulated charge, eq.~(\ref{eq:acc}), for MD profiles integrated up to $R_{\rm eff}$ (or similar, at least integrating over the first cationic and anionic adsorption monolayers), which can be tested directly against the effective charge obtained from fitting to the DH equation, as explained after eq.~(\ref{eq:fit}).  
	
	\section{Results and discussions}
	
	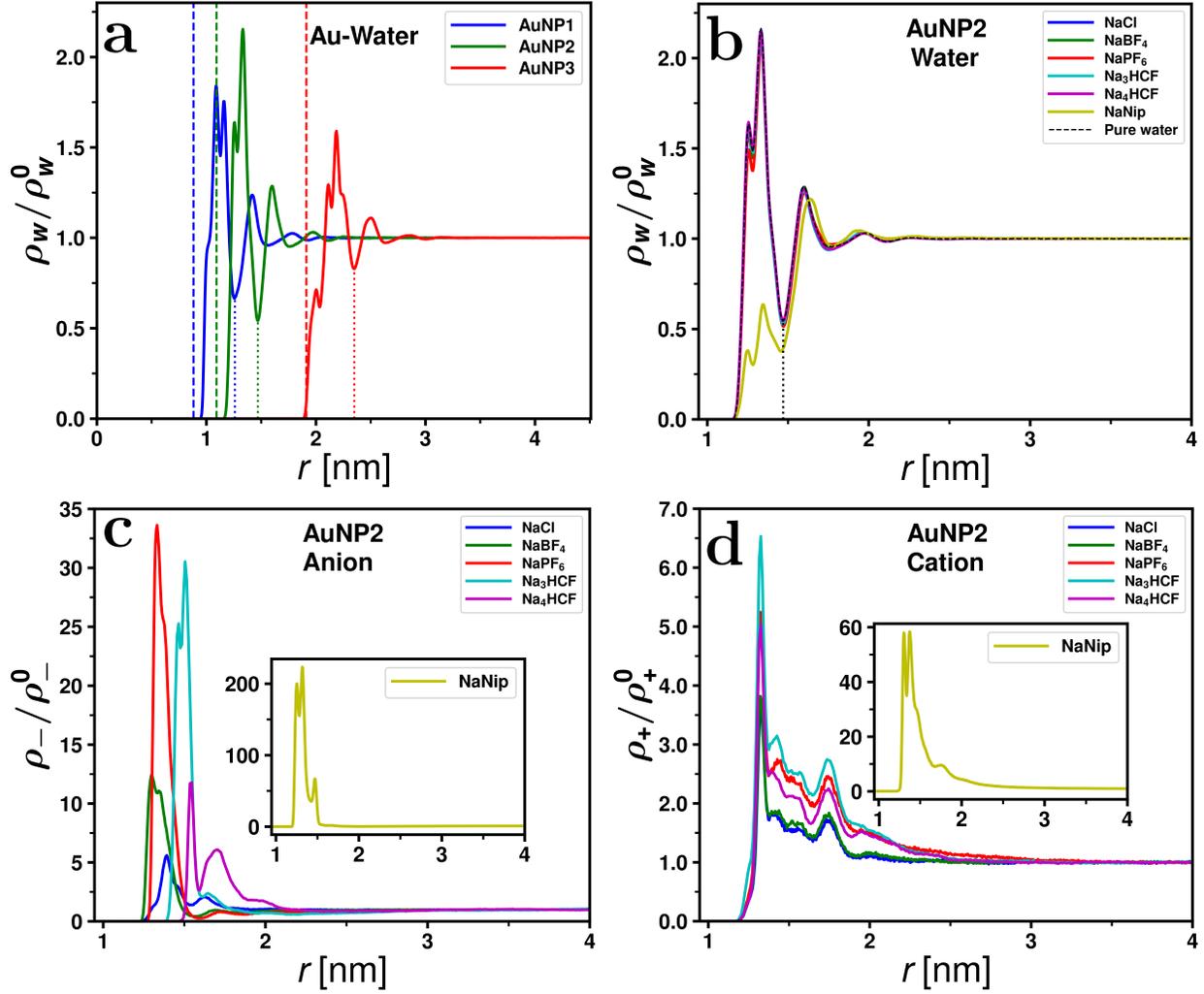
\begin{figure}[h!]
		\centering
		\includegraphics[width=1\linewidth]{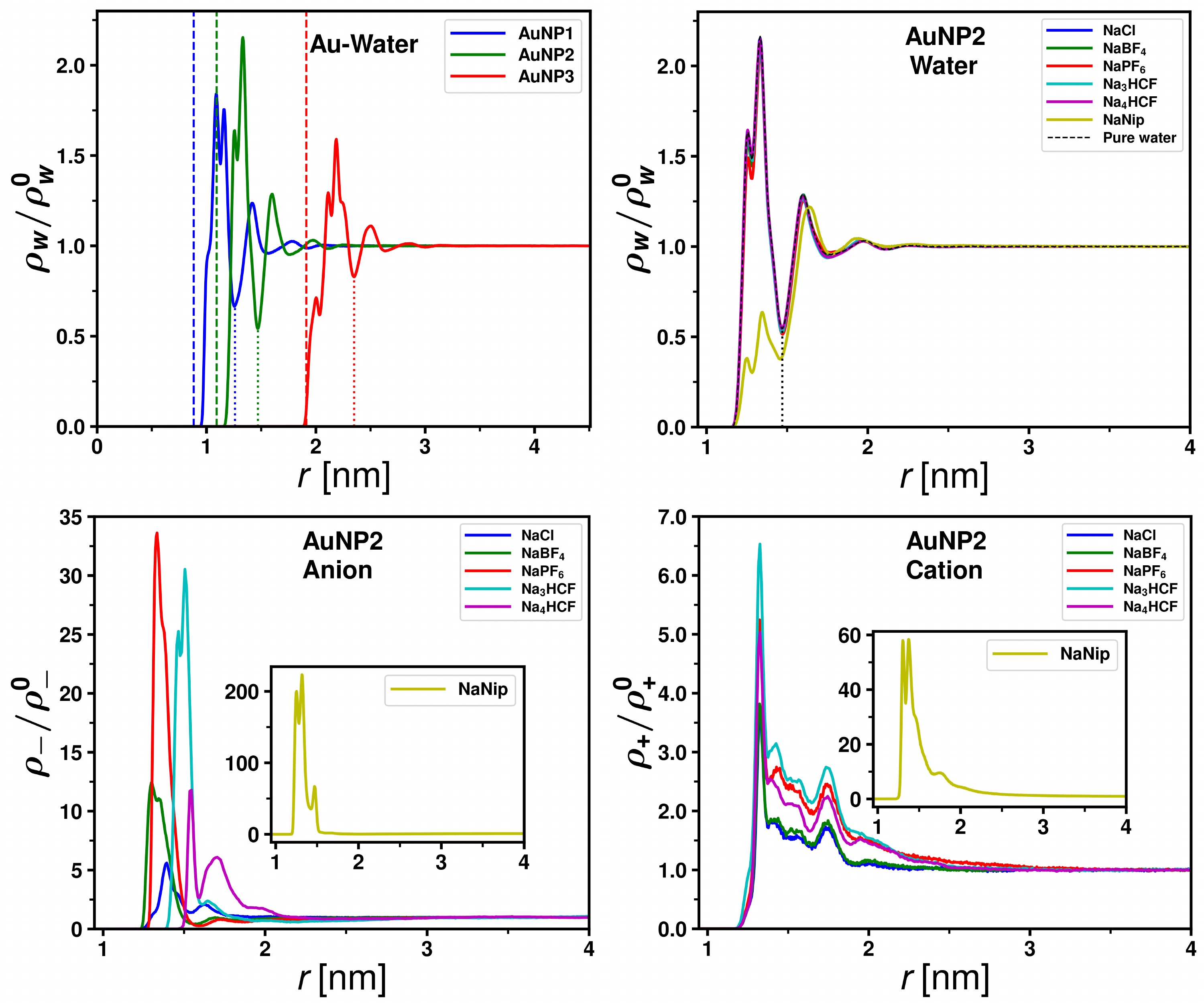}
		\put(-430,368){\textbf{\Huge a}} 
		\put(-195,365){\textbf{\Huge b}}
		\put(-430,175){\textbf{\Huge c}}
		\put(-195,168){\textbf{\Huge d}}
		\caption{Radial number density distributions $\rho_i(r)$ normalized by bulk density $\rho_i^0$ as a function of $r$ from the center of the AuNP. a) Water density profiles around the three different nanoparticles. The radii of the Gibbs dividing surface ($R_{\rm{GDS}}$) and effective surface ($R_{\rm{eff}}$) are shown by vertical dashed lines and vertical dotted lines, respectively. b) Water density profiles around AuNP2 with (solid lines) or without (dashed line) salt (see legend). c) Anion and d) cation density profiles around AuNP2 in the presence of different salts. The corresponding ionic strengths are given in Table~1.}
		\label{fig:rdf}
	\end{figure}
	
	\subsection{Water and Ion-specific Adsorption.}
	Figure~\ref{fig:rdf}a shows the normalized radial density distributions $\rho_{\rm w}(r)$ of water molecules surrounding the three different AuNPs in pure water. The water density profile displays two pronounced overall peaks for each of the AuNP systems, which is in agreement with the reported formation of water bilayer at gold surfaces.~\cite{Chang2008a,Geada2018a,Nadler2012a} The water bilayer is the energetically most favorable structure on close-packed hexagonal metal surfaces.\cite{Gross2020} More dense water adlayers are observed near the extended (111) facets, leading to large and narrow peaks close to the Gibbs dividing radius (eq.~(\ref{eq:GDS}), vertical dashed lines), followed by density minima beyond the first solvation shell. As argued in the Methods section, we defined the radius $R_{\rm eff}$ as the location of the global density minima following the first hydration shell. The obtained values are $R_{\rm eff}=$ 1.26, 1.47, and 2.35 nm for AuNP1, AuNP2, and AuNP3, respectively, are also summarized in Table~1. 
	
	The ion-specific effects on water adsorption on the AuNPs (taking AuNP2 as the reference system) are shown in Figure~\ref{fig:rdf}b. Except for the system with the Nip$^-$ anion, the ion adsorption plays a negligible role for the water radial distribution, particularly in the first solvation shell, and hence does not specifically change the local water structure compared with that of pure water. In contrast, for the Nip$^-$ anion-based system, the pure water distribution is substantially perturbed within the first solvation shell of the AuNP, \textit{i.e.}, the dense water layer at interfaces is broken down in the presence of NaNip, indicating large adsorption to the gold of the latter. However, there is a consistent local density minima at around 1.47 nm for water distributions in all studied AuNP2-based systems, which means that the location of $R_{\rm eff}$ (the presumed shear plane) of AuNP2 could be considered the same regardless of the anions.  
	
	Figure~\ref{fig:rdf}c displays the adsorption of anions on AuNP2, from which substantial ion-specific effects can be observed. Cl$^-$ gets least adsorbed in terms of the RDF peak height and Nip$^-$ is the most heavily adsorbed to the interface among all the studied anions. As can be seen in Figure~\ref{fig:rdf}d, the adsorption of cations (Na$^+$) follows the adsorption trends of the anions but the local binding is weaker as signified by significantly lower peak heights in the RDFs. In general, preferential adsorption of anions onto gold surfaces have been indeed usually observed when compared to the more `passive' cations, inferring mostly negative zeta-potentials as measured experimentally for bare gold nanoparticles~\cite{Dougherty2008}. 
	
	\begin{table}[h!]
		\centering
		\caption{Summary of the surface excess adsorption (over water) of ions and coordination numbers for the AuNP2 system. $N_{\pm}^{\rm ex}$ is the number of surface excess for cations and anions at the GDS radius $R_{\rm GDS}$, according to eq.~(2). $N_{\pm}$ are the coordination numbers of cations and anions, respectively, and $Z$ is the accumulated net charge at distance $R$, according to eqs.~(3) and (4), respectively.  Note: For the calculations of coordination numbers, the values of cutoff distance $R$ were set to $R_{\rm{eff}}$ for systems containing NaCl, NaBF$_4$, NaPF$_6$, and NaNip, while $R$ was set to a larger value indicating the global first minimum of the anion number density profile (in Figure 2c) for the Na$_3$HCF and Na$_4$HCF systems.} 
		\label{tblS:sum} %
		\begin{tabular}{ccccccccccccc}
			\hline 
			\rule{0pt}{4ex}Salt&$N_+^{\rm ex}$&$N_-^{\rm ex}$ &$N_+$&$N_-$ &$Z$ \\ [1.5ex] 
			\hline 
			\rule{0pt}{3ex} NaCl     & 0.87& 0.83& 0.77 & 1.15 & $-$0.38 \\
			\rule{0pt}{4ex} NaBF$_4$ & 1.28& 1.20& 0.75 & 2.76 & $-$2.01 \\
			\rule{0pt}{4ex} NaPF$_6$ & 3.24& 3.20& 1.01 & 5.41 & $-$4.40 \\
			\rule{0pt}{4ex} NaNip    & 29.0& 29.0& 12.0 & 30.9 & $-$18.9 \\
			\rule{0pt}{4ex} Na$_3$HCF& 3.70& 1.12& 3.75 & 2.44 & $-$3.57 \\
			\rule{0pt}{4ex} Na$_4$HCF& 2.56& 0.61& 3.00 & 1.10 & $-$1.40 \\
			[1ex]\hline 
		\end{tabular}\quad 
	\end{table}
	
	We have also calculated the excess adsorption numbers of ions with respect to the Gibbs dividing surface according to eq.~(2), see Table~\ref{tblS:sum}, which reflect well the trends observed in Figure~\ref{fig:rdf}. However, the order does not simply follow the peak heights in the RDFs since the adsorption is an integrated quantity and thus also the position and width of the peak matter. The anionic {\it charge} adsorption ($= |z_- N_-^{\rm ex}|$ ) follows an overall order of  Cl$^-$ $<$ BF$_4$$^-$ $<$  HCF$^{4-}$ $<$ PF$_6$$^-$  $\lesssim$ HCF$^{3-}$   $<$ Nip$^-$. Note that due to electrostatic cooperativity, the total salt adsorption (the sum of cationic and anionic number adsorptions) follows the same order.
	
	\begin{figure}[h!]
		\centering
		\includegraphics[width=1.\linewidth]{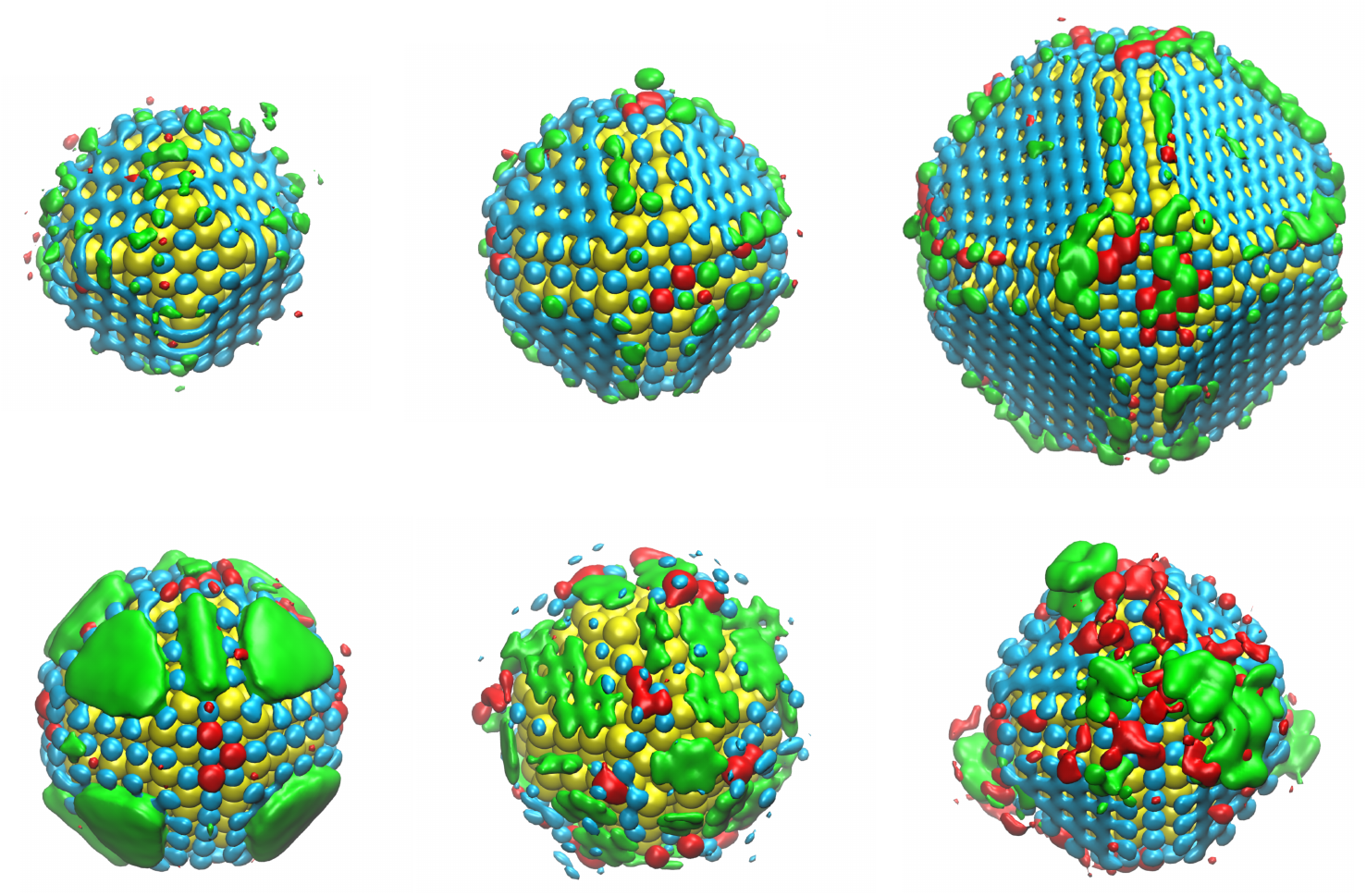}
		\put(-460,280){\textbf{\Huge a}}
		\put(-310,280){\textbf{\Huge b}}
		\put(-160,280){\textbf{\Huge c}}
		\put(-460,120){\textbf{\Huge d}} 
		\put(-310,120){\textbf{\Huge e}}
		\put(-150,120){\textbf{\Huge f}}
		\caption{Spatial (3D-resolved) density distribution profile of water and ions shown by iso-density surfaces in different colors. NaCl solutions for a) AuNP1, b) AuNP2, and c) AuNP3 systems. d) NaPF$_6$, e) NaNip and f) Na$_3$HCF solutions for AuNP2 system (red: cations, green: anions, light blue: water oxygens). The iso-density surface for water shows the density values of 60 nm$^{-3}$ and for cations and anions of 0.6 nm$^{-3}$ for all systems except NaNip, where 10 nm$^{-3}$ was used for the ions.}
		\label{fig:sdf}
	\end{figure}
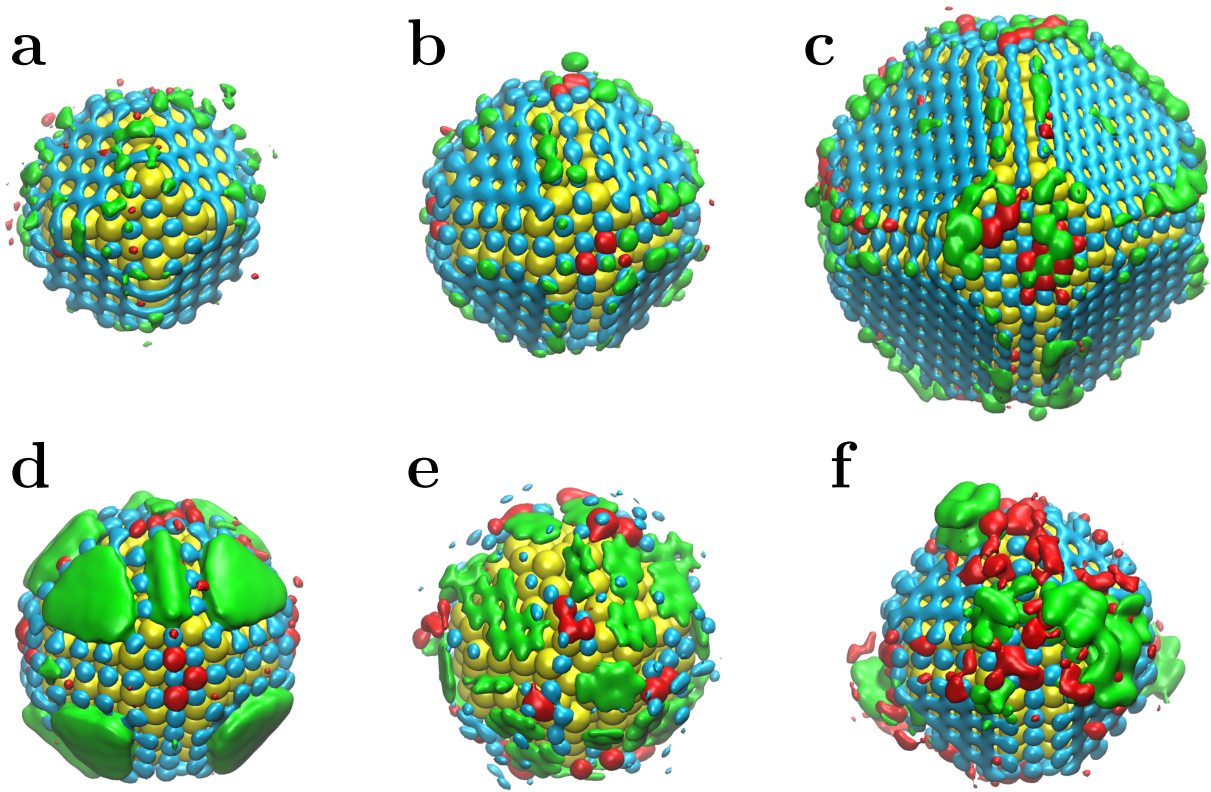
	
	\subsection{Three-dimensional (3D) Spatial Distribution of Water and Ions.}
	
	The spatial distribution functions (SDFs) were calculated with TRAVIS program~\cite{Brehm2011a} and the plots were created by VMD package~\cite{Humphrey1996}. The dense water layers observed in Figure~\ref{fig:rdf}a form well structured adlayers in space surrounding all three AuNPs as visualized in the spatial density distributions (Figure~\ref{fig:sdf}a-c). The formation of ordered, hexagonal water structure is observed on the extended planar (111) surfaces, while leaving some more disorder on the more `edgy' (100) and (110) facets. Water is indeed known to have an energetically favorable first hydration layer on the bare planar (111) gold surface.~\cite{Chang2008a,Geada2018a,Velez2014,Nadler2012a,Gim2019b} The spatial adsorptions of Na$^+$ and Cl$^-$ show very heterogeneous behavior with their preferential adsorption only on the disordered water regions (probably benefiting from broken hydrogen bonds, which may then favorably order in ionic hydration shells) and hardly on the (111) surfaces with the highly packed hexagonal water adlayers. 
	
	In addition to the NaCl systems, the (3D) spatially resolved adsorption structure of water and other salts are shown in Figure~\ref{fig:sdf}d-e and Figure~S2. Similar adlayer structures of water molecules as in Figure~\ref{fig:sdf}b are maintained for systems containing BF$_4$, PF$_6$, HCF$^{3-}$, or HCF$^{4-}$ anions. A quite heterogeneous adsorption of ions again is observed, especially for HCF$^{3-}$ and HCF$^{4-}$ anions where in addition an aggregation of ions is revealed, probably owing to the multivalency and the resulting larger charge--charge correlations. In contrast, the water spatial adsorption in the presence of Nip$^-$ anions shows a very different picture. In line with the radial distribution results, the water adlayers are substantially broken by the Nip$^-$ anions, which adsorb on the surface on average in more close contact, as compared to the other ions at similar ionic strength. Herein, an interesting adsorption structure of the Nip$^-$ anions is observed in Figure~\ref{fig:sdf}e, \textit{i.e.}, a part of Nip$^-$ anions are adsorbed flat on the gold (111) and (100) facets, while some anions are adsorbed on the edges at the narrow (110) facets by altering their rotations.  Another, apparently different adsorption mechanisms is exhibited by PF$_6^-$, where the adsorption is quite pronounced mostly on the flat (111) facets, but the first water layer stays more intact than for the case of Nip$^-$. 
	
	\begin{figure}[htb]
		\centering
		\scalebox{0.8}{
			\includegraphics[width=1\linewidth]{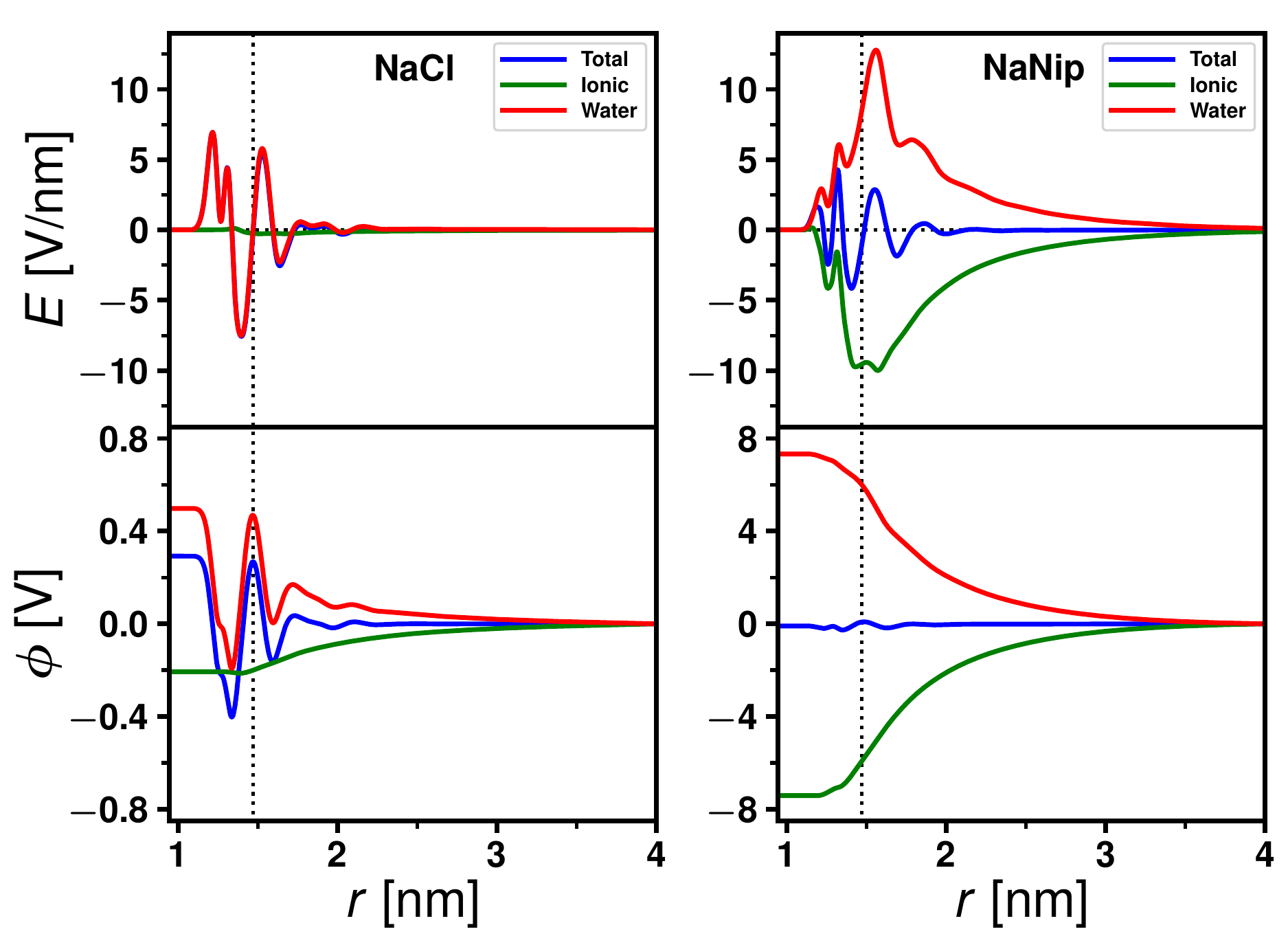}
			\put(-405,315){\textbf{\Huge a}} 
			\put(-180,312){\textbf{\Huge b}}
			\put(-405,172){\textbf{\Huge c}}
			\put(-25,168){\textbf{\Huge d}}
		}%
		\caption{Electrostatic properties of two representative salts for the AuNP2 system.  Electric fields $E(r)$ calculated by eq.~(\ref{eq:field}) for a) NaCl and b) NaNip solutions.  c)-d) Corresponding electrostatic potentials $\phi(r)$ obtained by eq.~(\ref{eq:phi}). The total value (blue line) for each of them is decomposed into two parts, \textit{i.e.}, ionic contribution (green line) and contribution from water molecules (red line). The vertical dotted lines represent the position of effective surface ($R_{\rm eff}$).}
		\label{fig:electros.}
	\end{figure}
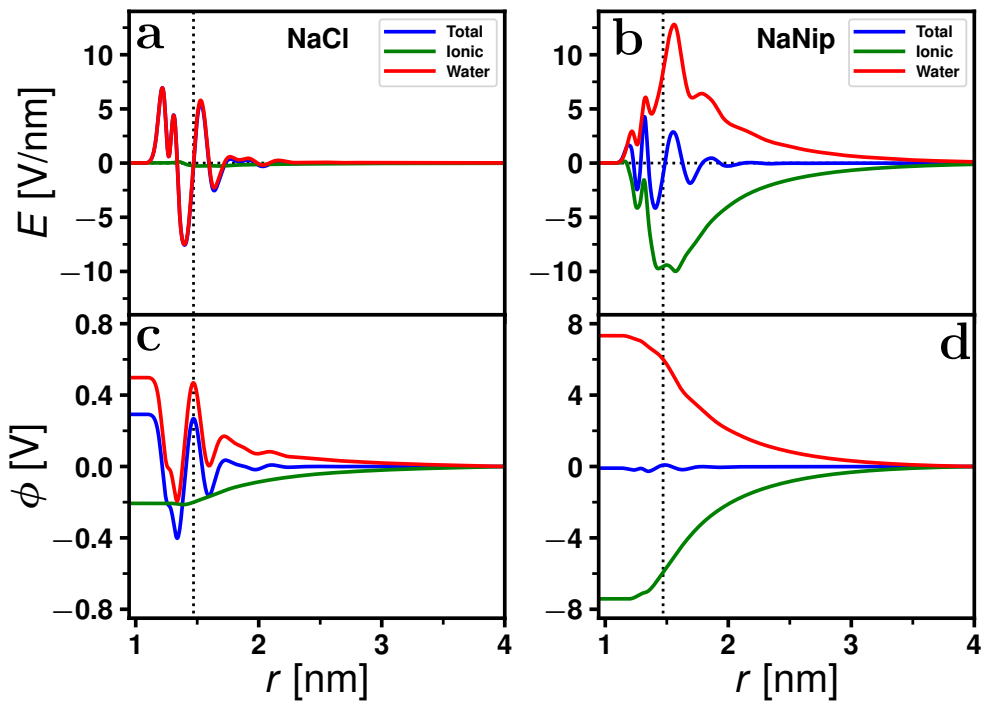 
	
	\subsection{Intermediate Discussion: Possible Effects of Ion-specific Adsorption on Functional Interfaces}
	
	The location and strength of ion binding onto an AuNP play an important role in its surface reaction processes for catalytic applications.\cite{enzyme,restructuring,Wu2012a,Herves,catalysis,synthesis,defect} The radial number density (in Figure 2) combined with the 3D spatial distribution (in Figure 3) show distinct ionic adsorptions. In particular, a simple anion as Cl$^-$ is known to preferentially adsorb, albeit relatively weakly to flat gold electrodes at zero potential.~\cite{Magnussen2002} Moreover, the observation of the strongest adsorption of Nip$^-$ at the interface is probably due to its aromatic character along with its platelike structure, being favorable for adsorbing at flat, relatively nonpolar surfaces. Nip$^-$ was found particularly strongly adsorbed with epitaxial matching on the (111) facets (see Figure~S3 in SI), owing to the soft epitaxial adsorption mechanism as proposed by Feng and co-workers, where the coordination of more polarizable atoms (such as C, N, O) in peptides or surfactants of multiple epitaxial sites favors the adsorption on gold (111) facets.~\cite{Feng2011b,Feng2012b} On the one hand, the hydrophobic nature of Nip$^-$ enables these anions to favorably release water molecules and to adsorb on AuNP surfaces. On the other hand, from an electronic structure perspective, it is well established that organic molecules with aromatic rings strongly physisorb on metal surfaces due to dispersion interactions, which in our case are approximately included in the Lennard-Jones parameters. In fact, density-functional theory calculations with comparable calculation settings suggest that the adsorption energy of a single benzene molecule is around 2.5 times stronger than the adsorption energy of a single water molecule on Au(111) (see Refs.~\citenum{Nadler2012a,CarrascoLiu2014}). Our results suggest that our simulations capture these effects, at least qualitatively.
	
	The micro-electrostatic structure right at the solid/electrolyte interface is determined by a complex interplay of interactions between solutes, solvent, electrolytes, and the surface, having itself an essential role in the electrochemical and catalytic properties of the system. It has been shown that the electronic structure of metals may not be greatly modified by the adsorption of a water bilayer, but the work function of many metals gets reduced.~\cite{defect,roy, Schnur_2009} Furthermore, the work function strongly depends on the orientation of the water molecules. For Au(111) and other metals, the work function changes by about 2 eV depending on the orientation of the hydrogen atoms with respect to the surface due to the opposite dipole orientation.~\cite{Gross2020,Schnur_2009}. In this regard, the presence of electrolytes which significantly modify the water structure at the interface will have substantial effects on electronic charge transfer. Our results show that this could be significant for NaNip and NaPF${_6}$ solutions. Note again that Nip$^-$ reduction at AuNPs by borohydride ions is an important model reaction in metal nanoparticle catalysis,~\cite{Herves}  and can be tuned by the electrostatic charge and surface potential of the nanoparticle.~\cite{roy} Local electrostatic features will be discussed in detail now in the following sections.
	
	\subsection{Water and Ionic Contributions to Electrostatic Properties.}
	
	We first perform calculations of electrostatic properties using an ``explicit water'' approach: here, the atomic partial charges of all species, including ions and water molecules, are taken into account for calculating the radial electrostatic properties. This approach  assumes the relative permittivity $\epsilon_{\rm{r}}$ = 1 due to the vacuum reference medium. The total electric field can be decomposed into two parts according to eq.~(\ref{eq:field}): the first part is the ionic contribution and the second contribution comes from the water molecules.  For the NaCl system, the decomposed electric field shows that the ionic contribution to the total one is negligible compared with that from the water molecules (Figure~\ref{fig:electros.}a). In the decomposed electrostatic potential calculated in eq.~(\ref{eq:phi}), the relative ionic contribution from the integration of the electric field is larger and negative, while water molecules control the short range, oscillatory structure of the total potential (Figure~\ref{fig:electros.}c). In the far field ($r\gtrsim 2$~nm) water and ionic contributions seem to cancel on this scale due to the large dielectric screening of water. 
	
	For the NaNip system, the ionic species substantially contribute to both the total electric field and potential (Figure~\ref{fig:electros.}b and d) as a consequence of the strong adsorption of Nip$^-$ at interfaces altering the local charge distributions in contrast to NaCl system (Figure~S4 in SI). The water-induced electric field and potential are increased obviously along with the ionic ones to eliminate the excessive field and potential.  In addition, the electric fields and potentials for the rest of the studied systems containing other anions are also calculated (Figure~S5 in SI). Generally, the potential induced by ions is negative and to large extent canceled out by the opposite water contribution at large distances ($>$2~nm). The resulting total potential is by the factor $1/\epsilon_\textrm{r}\approx 0.014$ (dielectric effect of water) smaller than the bare ionic potential.
	
	\begin{figure}[htb]
		\centering
		\scalebox{0.8}{
			\includegraphics[width=1.\linewidth]{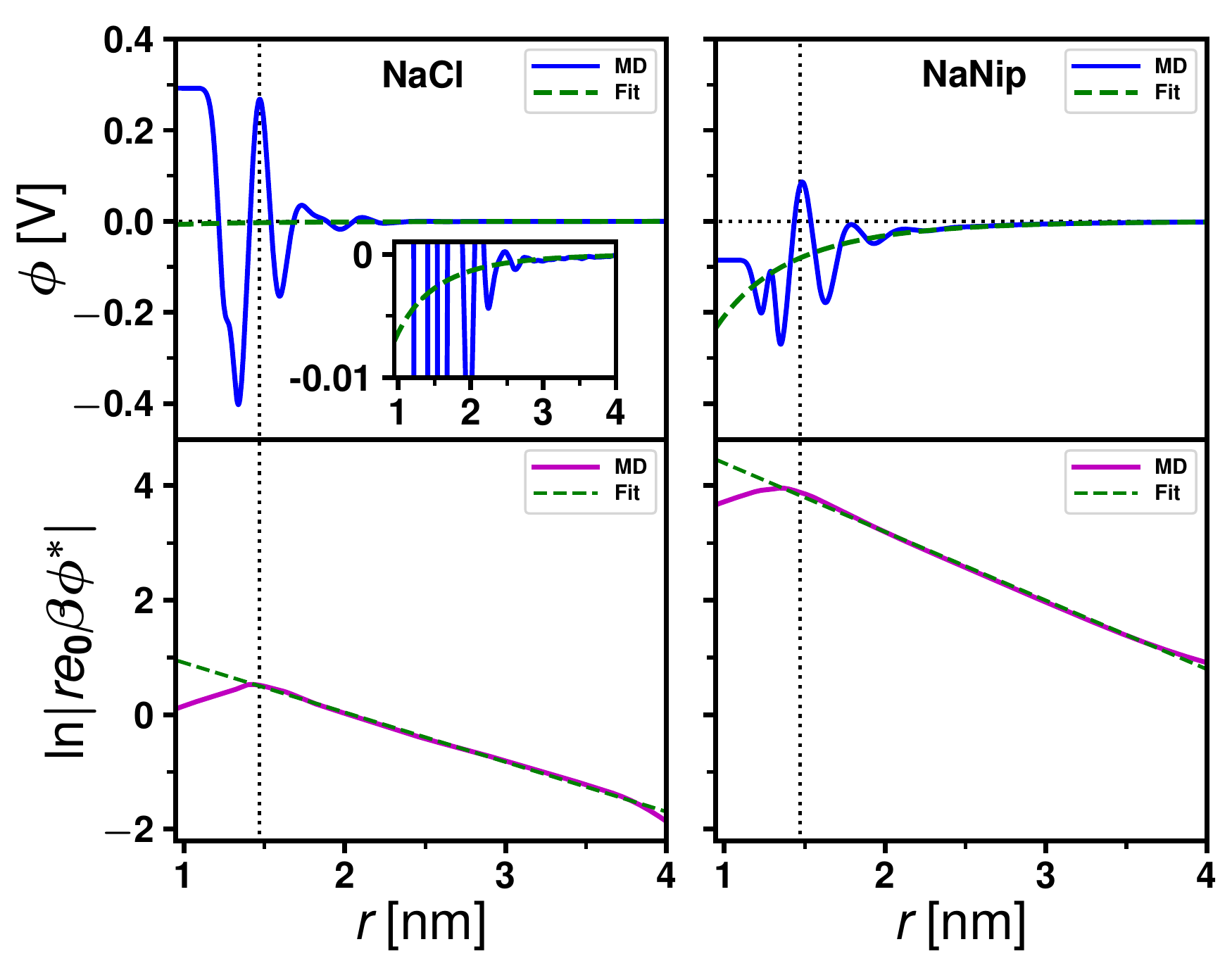}
			\put(-400,340){\textbf{\Huge a}} 
			\put(-195,335){\textbf{\Huge b}}
			\put(-400,185){\textbf{\Huge c}}
			\put(-120,180){\textbf{\Huge d}}
		}%
		\caption{Electrostatic potentials and DH fits for two representative salt systems. a)-b) Total electrostatic potentials obtained from MD simulations using the explicit-water approach (blue lines) and the DH fits (green dashed lines) for NaCl and NaNip solutions. c)-d) Logarithmic plots of the rescaled electrostatic potential obtained from the implicit-water approach (excluding charge density distributions of water and assuming $\epsilon_r$ = 71), shown by purple lines. The dashed lines are the linear fit to the far-field regime. The vertical dotted lines represent the position of the effective surface ($R_{\rm eff}$).}
		\label{fig:DH}
	\end{figure}
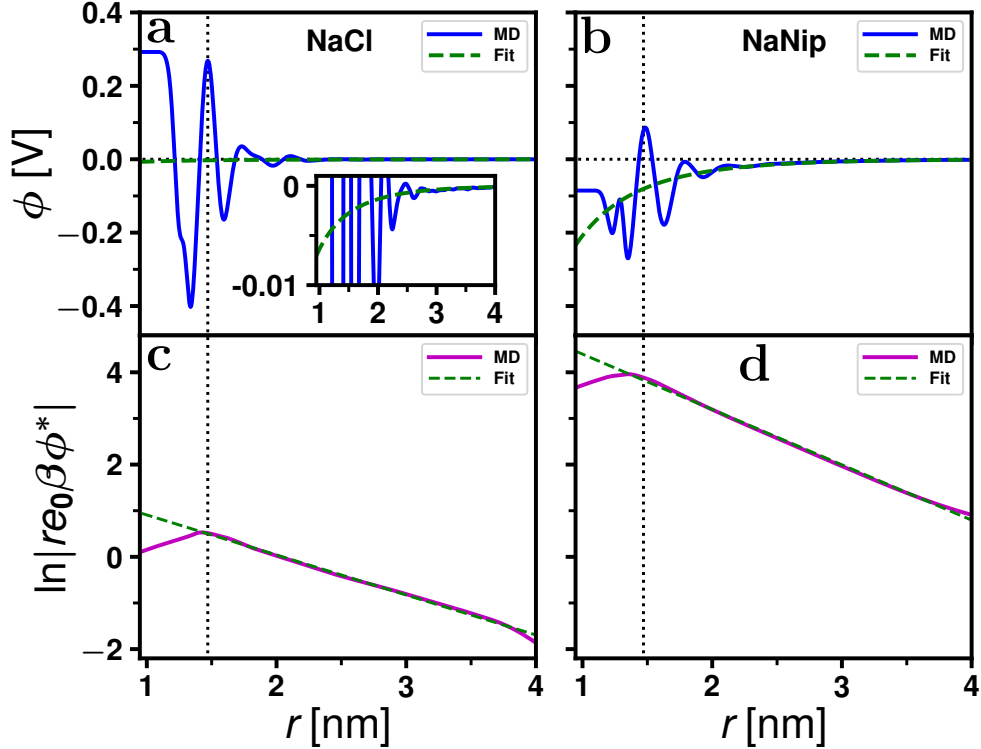
	
	\begin{table}[h!]
		\centering
		\caption{Summarized properties of the studied solutions for the AuNP2 systems. $I$ is the  ionic strength (evaluated close to the cell edges, see text), $\lambda_{\rm{D}}$ and $\lambda_{\rm{D}}^{\rm{fit}}$ are the theoretical Debye length calculated from the ionic strength ($I$) and the Debye length obtained from the DH fit to the implicit-solvent charge distributions, respectively. $Z_{\rm{eff}}$ and $\sigma_{\rm{eff}} = Z_{\rm eff}/(4\pi R_{\rm eff}^2)$ are the effective surface charge and the effective charge density at $R_{\rm{eff}}$ from the DH fits, respectively. $~\phi_{\rm{eff}}^{\rm{DH}}$ and $\phi_{\rm{eff}}^{\rm{MD}}$ are the electrostatic potential from the DH fit and from the total MD charge distributions explicitly calculated at $R_{\rm{eff}}= 1.47$~nm.} 
		\label{tbl:sum}
		\begin{tabular}{ccccccccccccc}
			\hline 
			\rule{0pt}{4ex}Salt&$I$ [mM] & $\lambda_{\rm{D}}$ [nm] & $\lambda_{\rm{D}}^{\rm{fit}}$ [nm]& $Z_{\rm{eff}}$ [$e_0$] & $\sigma_{\rm{eff}}$ [$e_0$/nm$^2$] & $\phi_{\rm{eff}}^{\rm{DH}}$ [mV]& $\phi_{\rm{eff}}^{\rm{MD}}$ [mV] \\ [1.5ex] 
			\hline 
			\rule{0pt}{3ex} NaCl     & 131(1)& 0.80 & 1.13 & $-$0.5& $-$0.02& $-$3  & 268 \\
			\rule{0pt}{4ex} NaBF$_4$ & 119(2)& 0.84 & 0.78 & $-$1.6& $-$0.06& $-$8  & 267 \\
			\rule{0pt}{4ex} NaPF$_6$ & 113(1)& 0.86 & 0.80 & $-$3.9& $-$0.14& $-$19 & 248 \\
			\rule{0pt}{4ex} NaNip    & 93(3) & 0.95 & 0.83 & $-$16.1& $-$0.59& $-$81 &  81 \\
			\rule{0pt}{4ex} Na$_3$HCF& 249(5)& 0.58 & 0.62 & $-$5.3& $-$0.20& $-$22 & 249 \\
			\rule{0pt}{4ex} Na$_4$HCF& 287(13)& 0.54 & 0.71 & $-$3.2& $-$0.12& $-$15 & 265 \\
			[1ex]\hline 
		\end{tabular}\quad 
	\end{table}
	
	\subsection{`Real' and Effective Surface Potentials and Charges}
	
	From the explicitly integrated MD charge profiles we can directly read off the realized surface potentials for various surface definitions. The values at the Gibbs dividing radius are given in Table~S1 in the SI. The electrostatic potentials $\phi_{\rm eff}^{\rm MD}$ at $R_{\rm eff}$, are summarized in Table~2 for AuNP2 for all salts. Due to the highly oscillatory nature of the electrostatic potential close to the interface (see the potentials near $R_{\rm eff}$ in Figure~\ref{fig:DH}a and b), the values of $\phi_{\rm eff}^{\rm MD}$ are very sensitive to the exact choice of the interface radius, which is ill-defined at these small scales.  For applications on larger, colloidal scales, the effective surface potential in the DH framework (apparent value from a far-field view) is more relevant, as we discuss in the following. 
	
	The electrostatic potential profile can be mapped onto a DH type of potential in the far field as described by eq.~(\ref{eq:DH}). The fit on the total potential using eq.~(\ref{eq:fit}) then yields the effective charge $Z_{\rm eff}$ and potential $\phi_{\rm eff}^{\rm DH}$ at a given radius $R_{\rm eff}$. For this, the total electrostatic potentials (same as the total ones in Figure~\ref{fig:electros.}c and d) obtained from the MD simulations are shown in Figure~\ref{fig:DH}a and b as a function of radial distance for two representative systems with AuNP2, and fitted by the DH equation~(\ref{eq:DH}).  However, as we can see from Figure~S6 in the SI, which is the logarithmic version of Figure~5a and b, the potential exhibits profound fluctuations due to the explicit water molecules.  
	
	An elegant solution to this problem is the ``implicit water'' approach~\cite{Xu2017c,Nikam2018}, in which the partial charges of water molecules are excluded from eq.~(\ref{eq:Poisson}) when calculating the electrostatic potential, and instead the relative permittivity of the aqueous medium is set to $\epsilon_\textrm{r}=71$. The potentials calculated in this way, labeled as $\phi^*$ (Figure~\ref{fig:DH}c and d), were fitted in the far field by the DH model based on eq.~(\ref{eq:fit}). It is clearly seen that the rescaled potentials could be well fitted to linear functions for both systems. The Debye lengths $\lambda_{\rm{D}}^{\rm{fit}}$ (= $1/\kappa^{\rm{fit}}$) obtained from the fits have similar values as the theoretical values at the corresponding `bulk' ionic strengths, established far from the NP in the simulation cell, see Table 2. Note that very close to $R_{\rm eff}=1.47$~nm, the potential deviates from linearity, justifying our definition of $R_{\rm eff}$, which separates the diffusive double layer (linear behavior) from more nonlinear regimes, in which typically the ions are more strongly correlated. The effective charge and potential at the given radius $R_{\rm eff}$ are determined from fitting eq.~(\ref{eq:fit}). The overall DH potential as a function of radial distance is then obtained as shown in Figure~\ref{fig:DH} by green dashed lines; see also Figure~S7 in the SI for other salts.   
	
	The effective charges $Z_{\rm{eff}}$ and effective potentials ($\phi_{\rm{eff}}^{\rm{DH}}$) for all salts at AuNP2 are summarized in Table~\ref{tbl:sum}.  The overall effective charges and potentials are all negative. The effective charge is smallest for the  NaCl system with a value of $-$0.5 $e_0$. For the Na$_3$HCF system, a negative charge with a value of $-$5.3 $e_0$ is by one order of magnitude larger than that of the NaCl system. This difference comes from the stronger anionic adsorption of Na$_3$HCF than NaCl. Furthermore, the NaNip system has the largest absolute value of the effective charge ($-$16.1 $e_0$), as is understandable from the strongest charge adsorption of anions. In general, the values of the effective charge at $R_{\rm eff}$  are ranging from $-$1.6 to $-$3.9 $e_0$ for the remaining studied systems with AuNP2, and the order is consistent with their total salt adsorption trends, see again in Table~\ref{tblS:sum}. 
	
	We also calculated effective nanoparticle charges from the accumulated charges $Z$ according to eq.~(\ref{eq:acc}) up to a radial distance $R$. The values for $R$ were set to $R_{\rm{eff}}$ for systems containing NaCl, NaBF$_4$, NaPF$_6$, and NaNip, whereas for the Na$_3$HCF and Na$_4$HCF systems, $R$ was fixed to a slightly larger value, which matches the first global minimum of the anion number density profile (in Figure 2c) in order to integrate over the entire adsorbed molecule. Comparing the values of $Z$ (Table~\ref{tblS:sum}) and $Z_{\rm eff}$ (Table~\ref{tbl:sum}), we find them indeed following the same anionic series, and are almost quantitative in some cases. Hence, we obtain consistent values from two independent approaches, from a far-field DH fit and from the accumulated net charge (both with respect to essentially the same effective surface definition). 
	
	\begin{figure}[htb]
		\centering
		\includegraphics[width=0.6\linewidth]{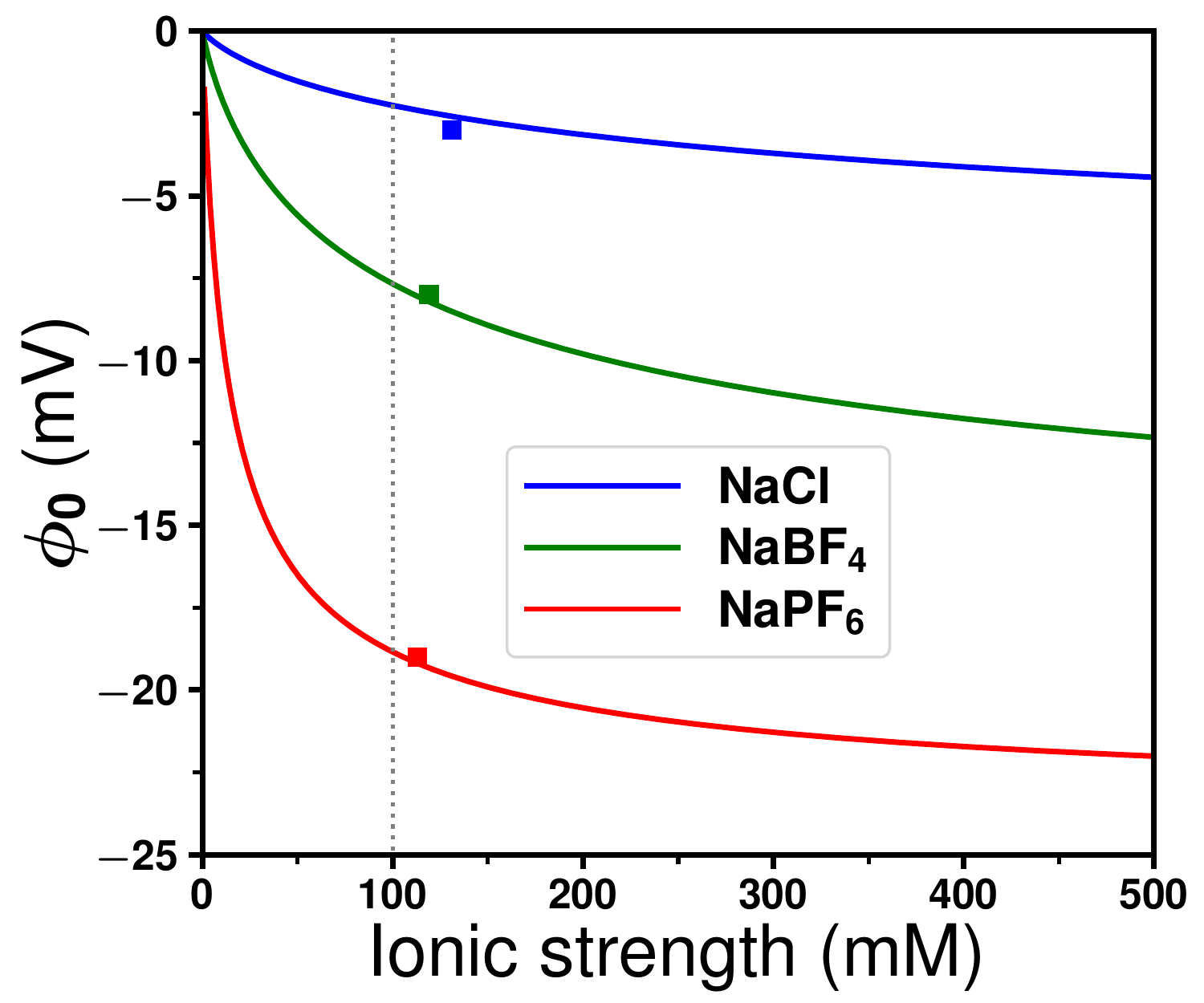}
		\caption{Effective surface potential at $R_{\rm{eff}}$ as a function of ionic strength obtained by eq.~(\ref{eq:surf-pot}). The squared points represent the values in our simulations at the studied ionic strengths.}
		\label{figS:phi-vs-c}
	\end{figure}
	
	The calculated overall effective potentials (expected to be close to zeta-potentials) ranging from $-$3 to $-$81 mV are in line with typical ranges measured in experiments.~\cite{Merk2014b,Dougherty2008} Again, the NaNip system yields the largest negative effective potential ($-$81 mV) at $R_{\rm eff}$. Note, the ionic strengths are slightly different among the various salts, which makes the comparison of effective potentials not so straightforward. However, combined with the Adsorption-Grahame eq.~(\ref{eq:surf-pot}) and the DH potential obtained from our simulations at specific ionic strengths, the effective potential can be extrapolated to arbitrary ionic strengths. The calculated specific anion binding affinities ($\varepsilon_-$) for Cl$^-$, BF$_4$$^-$, PF$_6$$^-$ are $-$1.9, $-$4.5, $-$8.7 kJ/mol, respectively. By using these values, the effective surface potential for the corresponding systems can be plotted versus ionic strength as shown in Figure~\ref{figS:phi-vs-c}. The predicted effective potentials feature all the same monotonic decreasing behavior with increasing ionic strength. With such a prediction at hand,  we can consistently compare the potentials at a certain given ionic strength, such as 100 mM (vertical dotted line in Figure~6). Given the values in Table~\ref{tbl:sum} and considering the uncertainty of the deviation from the ideal DH model, the absolute value of the effective potential has an ionic ordering given by Cl$^-$ $<$ BF$_4$$^-$ $<$  HCF$^{4-}$ $\lesssim$ PF$_6$$^-$ $\lesssim$ HCF$^{3-}$ $\ll$ Nip$^-$, which again correlates well with the specific anionic (charge) adsorptions. 
	
	\subsection{Ion-specific Effects on Nanoparticle Stabilization}
	The surface potential plays a critical role in the colloidal dispersion stability of nanoparticles in solutions, which can be explained in the framework of the  Derjaguin--Landau--Verwey--Overbeek (DLVO) theory.~\cite{Hunter,deraguin1941,Verwey1948} Within this approach, the interaction between two charged nanoparticles is the sum of van der Waals (vdW) and electrostatic (double layer) interactions. The vdW interaction is attractive, depends only on the materials involved (i.e., gold nanoparticle (solid) and water (solvent)), and for two spherical particles with radius $R$ scales with the center-to-center distance ($r$) as $V_{\rm{vdW}}(r)\propto-1/(r-2R)$ for small distances. The electrostatic interaction, on the other hand, depends on the effective charge and ionic screening, $V_{\rm{el}}(r)\propto Z_{\rm{eff}}^2 \exp{(-\kappa r)}/r$, which, after expressing it in terms of the surface potential, becomes $V_{\rm{el}}(r)\propto \phi_0^2 \exp{(-\kappa r)}/r$. According to this picture, a dispersion solution of golden nanoparticles is not stable at very low electrolyte concentrations, since the surface potential $\phi_0$ is too low (see Figure~\ref{figS:phi-vs-c}) to compensate for the vdW attraction. Increasing the ionic strength and with that the surface potential of AuNPs, stabilizes the suspension. But only up to a certain point. Namely, $\phi_0$ saturates at high $I$ (Figure~\ref{figS:phi-vs-c}), whereas the increasing exponential screening factor $\exp{(-\kappa r)}/r$ reduces the repulsive $V_{\rm{el}}$, which eventually enables the vdW contribution to prevail again. Thus, each electrolyte type features a certain intermediate range of concentrations---in turn, resulting into a range of surface potentials---where the AuNP suspension is stable. Such a  non-monotonic stabilization behavior was indeed observed experimentally for AuNPs in monovalent salt concentrations of various types.~\cite{Merk2014b} However, the surface (charge) properties of the AuNPs in the experiments are not exactly known, so a direct comparison to our study cannot be performed.
	
	In addition to the above discussion regarding the effect of the electrostatic potential on stability, experimental studies have also demonstrated that colloidal stabilization and dispersion of laser-ablated NPs depend on the anionic species present in solution, following a Hofmeister-like series~\cite{Merk2014b}. In particular, chaotropic ions such as SCN$^{-}$, Br$^{-}$, and I$^{-}$ tend to stabilize the bare NPs while ions such as F$^{-}$ and SO${_4}^{2-}$ tend to do the opposite. Ions such as Cl$^{-}$ and NO${_3}^{-}$ exhibit an intermediate effect.  An inversion of the Hofmeister series has also been observed by switching the surface nature of colloidal particles from hydrophobic to hydrophilic~\cite{peula2010inversion,morag2013governing,oncsik2016charging}. With the exception of Cl${^-}$, the electrolytes studied in our work have not been investigated experimentally in this context to the best of our knowledge, making a direct comparison not possible.
	
	Our results can, nevertheless, bring insight to ion-specific effects regarding the stabilization of gold nanoparticles and its relation to the Hofmeister series. For example, ion-specific effects on colloidal stability can be correlated to the polarizability and hydration number of each specific ion, yielding a characteristic ranking that follows a Hofmeister-like series.~\cite{Merk2014b} The estimation of the polarizabilities and hydration number of the ionic species in our study is not easy at best and beyond the scope of our study. Fortunately, experiments have also suggested that the Hofmeister-like series in AuNPs can also be understood in terms of accumulation and depletion close to the interface of chaotropic and kosmotropic ions, respectively. Our results show a distinctive facet selectivity that is ion specific. In particular, BF$^{4-}$, PF$^{6-}$, and Nip$^{-}$ adsorb strongly on the extended flat (111) facets of the NPs, suggesting a stabilizing behavior in the NPs formation such as the chaotropic ions mentioned above. On the other hand, Cl$^{-}$ and HCF$^{3-}$ are present on the less extended (100) and (110) facets, which suggest an intermediate or destabilizing effect in NPs formation, closer to the behavior of kosmotropic ions. A more in-depth investigation of these effects would be worth pursuing by modeling specific ions present in AuNPs formation where experimental information is available, including polarizability effects on the NPs and the ionic species themselves.
	
	\section{Conclusion}
	
	In summary, we have studied the solvation structure, ion-specific adsorption, and electrostatic surface properties of bare gold nanoparticles for various anions in aqueous electrolytes (sodium salts) using classical molecular simulations. Although experimental studies typically consider functionalized nanoparticles, either by purpose or for synthesis reasons, our study on the simple bare nanoparticles sheds light on various fundamental aspects of ionic adsorption on metal nanoparticles.  The solvation and double-layer structure and calculated (real and effective) surface potentials exhibit strong ion-specific (Hofmeister-like) effects, which can be traced back to complex heterogeneous adsorption structures of the ions, which vary significantly in complexity in terms of charge densities and geometries. In particular, we observed high facet selectivity, for example, the rather disk-like Nip$^-$ anion adsorbed strongly on the flat and extended (111) facets of the nanoparticles, penetrating and disarranging the well-ordered first hydration layer more than the other ions, and leading to the highest observed effective surface potential of about $-$81~mV at bulk concentrations of about 100 mM.   
	
	The near-field electrostatic potentials right at the solid/electrolyte interface are very important for electrochemical and catalytic properties of the system.\cite{synthesis,defect,roy} We have demonstrated that these local interface potentials are substantially modified upon the adsorption of ions.  Moreover, since the work function for electrons strongly depends also on the orientation of the water molecules,~\cite{Schnur_2009,Gross2020} the presence of electrolytes which significantly modify the water structure at the interface will have substantial effects on electronic charge transfer.  Our results show that this could be highly significant for strongly adsorbing molecular ions, such as Nip$^-$. 
	
	In the far-field, the ion-specific electrostatic potentials determine the larger-scale response and properties of the nanoparticles, e.g., their electrophoretic mobility and their interaction with other charged macromolecules in solution. For example, strong Hofmeister effects have been observed in the stability of gold nanoparticle dispersion upon adding salt, specifically for various anions,\cite{Merk2014b} for which a quantitative microscopic interpretation is still lacking. There, however, oxidation effects upon laser ablation on the gold may play a substantial role. Hence, further simulation work on AuNPs could study the effects of laser-induced partial oxidation of surface Au atoms,\cite{Merk2014b} but also the effects of defects,\cite{defect} surface reconstruction,~\cite{restructuring} and functionalization\cite{Heinz2017} on anionic adsorption and the NP's electrostatic properties. In addition, explicit consideration of metal polarizability in modeling studies could be important. NP polarizability has a known influence on the ionic distributions in the solution.~\cite{Petersen2018d} In particular, NP surface edges and defects have certainly a different electrostatic polarizability behavior than bulk systems or simple planar surfaces, and therefore the explicit inclusion of dielectric/electronic polarizability in MD simulations, e.g., by classical dipoles or oscillators~\cite{Geada2018a,Perfilieva2019a}, or a continuum capacitance--polarizability force field~\cite{Li2015b} should be worth for future studies on NPs in electrolyte solutions. 
	
	\section{Associated content}
	\subsection{Supporting Information}
	
	Radial distributions of gold atoms for the three studied AuNPs; representative snapshots of Nip$^-$ anion adsorption on the (111) facet of AuNP2; spatial density distribution and electrostatic properties for other studied AuNP2 systems; decomposed charge density distributions; electrostatic properties at the Gibbs dividing surface for the AuNP2 systems.
	
	\begin{acknowledgement}
		This project has received funding from the European Research Council (ERC) under the European Union's Horizon 2020 research and innovation programme (Grant Agreement 646659-NANOREACTOR). M.K. acknowledges the financial	support from the Slovenian Research Agency (Contracts P1-0055 and J1-1701). The authors acknowledge support by the state of Baden-W{\"u}rttemberg through bwHPC and the German Research Foundation (DFG) through grant no INST 39/963-1 FUGG (bwForCluster NEMO).
	\end{acknowledgement}
	
	\bibliography{refs}
	
	\clearpage
	\section{For Table of Contents Only}
	\begin{figure}
		\centering
		\includegraphics[width=1.\linewidth]{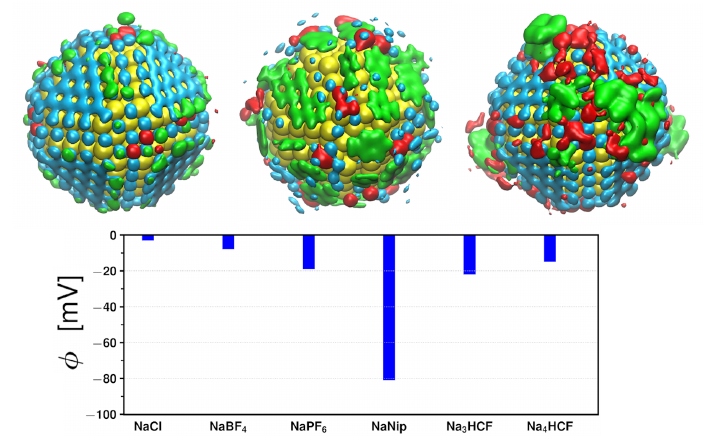}
	\end{figure}
\end{document}